\documentclass{article}
\usepackage[utf8]{inputenc}
\usepackage{amsmath}
\usepackage{amsfonts}
\usepackage[a4paper, total={6.5in, 9in}]{geometry}
\usepackage{graphicx}
\usepackage{color}

\usepackage[leftcaption]{sidecap}

\usepackage{amsmath}
\usepackage{amssymb}

\usepackage[numbers]{natbib}
\title{A note on the unbiased estimation of mutual information} 

\author{Jake Witter\thanks{jake.witter@bristol.ac.uk}{} and Conor Houghton\thanks{conor.houghton@bristol.ac.uk}\\Department of Computer Science, University of Bristol, England}

\begin{document}
\maketitle

\begin{abstract}

Estimators for mutual information are typically biased. However, in
the case of the Kozachenko-Leonenko estimator for metric spaces, a
type of nearest neighbour estimator, it is possible to calculate the
bias explicitly.

\end{abstract}

\subsection*{Introduction}

Consider the problem of calculating the mutual information $I(X,Y)$
between a discrete random variable $X$ and a variable $Y$ which takes
its value in a metric space $\mathcal{Y}$. This is a common situation,
$X$ can represent a set of labels while $Y$ represents a corresponding
space of outcomes which is high-dimensional or does not have a
convenient set of co\"{o}rdinates. Mutual information estimates for
high-dimensional data are important in biomedical science
\cite{MaesEtAl1997}, in blind source separation
\cite{BellSejnowski1995,HyvarinenEtAl2000}, information bottleneck
\cite{TishbyEtAl2000}, the analysis of deep learning networks
\cite{TishbyZaslavsky2015} and elsewhere and by considering mutual
information estimators which use only metric information it is
possible to avoid some of the difficulties that high-dimensionality
present.

In \cite{Houghton2015} a Kozachenko–Leonenko estimator
\cite{KozachenkoLeonenko1987, Victor2002, KraskovEtAl2004} is derived
for this situation. As is common with estimators for mutual
information this estimator has a bias; here this bias is calculated
exactly to give an unbiased estimator. This calculation closely
follows the calculation provided in \cite{Houghton2019} for a similar
estimator for the mutual information between two random variables
which both take values in a metric space.

\begin{SCfigure}[][bh]
\includegraphics[width=0.45\textwidth]{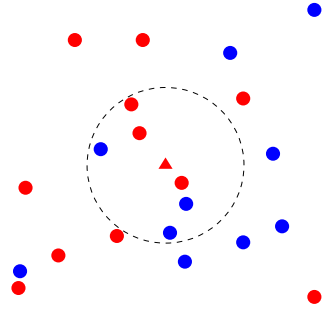}
\caption{\textbf{The calculation of} $I_0$. The circles and
  triangle are data points and red and blue represent two labels. The
  dashed line is the ball around the `seed' point in the center marked by a
  triangle $\color{red}\blacktriangle\color{black}$. Here $h=7$ so the
  ball has been expanded until it includes seven points. It contains
  four red points, the colour of the central point, so
  $h_y(\color{red}\blacktriangle\color{black})=4$. For illustration
  the points have been drawn in a two-dimensional space, but this can
  be any metric space. \label{fig:dots}}
\end{SCfigure}

When calculated the mutual information the data will be $(x,y)$ pairs:
\begin{equation}
  \{(x_1,y_1),(x_2,y_2),\ldots,(x_n,y_n)\}
\end{equation}
where $n$ is the number of data points. The estimator for mutual
information is a kind of nearest-neighbours or kernel density
estimator \cite{MoonEtAl1995,TobinHoughton2013}. It starts with a
smoothing parameter, an integer $h$. For each outcome $(x_i,y_i)$ a
ball is formed of the $h$ points, including $y_i$ itself, which are
closest to $y_i$ in $\mathcal{Y}$. $(x_i,y_i)$ will be called the seed
point for this ball containing $h$ points. Since the data are $(x,y)$ pairs,
each of these $h$ closest points, $y_j$, will have an associated label
$x_j$. To estimate the mutual information the number, $h_y(i)$, of
these $h$ points that has the same label as the seed is calculated:
\begin{equation}
  h_y(i)=\#\{x_j=x_i\mbox{ and }y_j\mbox{ is one of the }h\mbox{ closest points to }y_i\}
\end{equation}
The estimated mutual information is now
\begin{equation}
I_0=\frac{1}{n}\sum_{i=1}^n \log_2{\frac{n_xh_x(i)}{h}}  
\end{equation}
where $n_x$ is the number of different labels, that is, the number of
different possible outcomes for $X$. This is illustrated in
Fig.~\ref{fig:dots}.

This estimator $I_0$ is calculated using the metric structure on
$\mathcal{Y}$; it depends only on the matrix of distances between the
$y_i$ points and on their corresponding labels $x_i$. This distance
matrix in turn is used to work out the near $h$ points to each of the
$y_i$ in turn; the algorithm is not computationally fast because the
points need to be sorted in distance order. However, since it depends
only on the metric it works well for high-dimensional data.

\subsection*{Result}

This estimated is biased because it gives a non-zero value even if the
$X$ and $Y$ are independent. However, one advantage of the
Kozachenko–Leonenko is that bias can be calculated exactly. The bias is
given by
\begin{equation}
  I_b=\sum^{h}_{r=1} \mathbb{P}(h_y = r) \log_2  {\frac{n_x r}{h}}
\end{equation}
where $\mathbb{P}(h_y = r)$ is the probability $h_y$ is $r$ if $X$ and
$Y$ are independent.

Now
\begin{equation}
  \mathbb{P}(h_y = r)=\sum_{c=1}^{n_x} \mathbb{P}(x=x_c)\mathbb{P}(h_y = r|x=x_c)
\end{equation}
where the sum is over the possible labels $x_c$ of $X$. Let $n_c$ be
the number of data points with label $x_c$, so
\begin{equation}
  n=\sum_{c=1}^{n_x}n_c
\end{equation}
then the bias for the data set is
\begin{equation}
  \mathbb{P}(h_y = r)=\sum_{c=1}^{n_x} \frac{n_c}{n}\mathbb{P}(h_y = r|x=x_c)
\end{equation}
Calculating $\mathbb{P}(h_y=r|x=x_c)$ is an urn
problem. $\mathbb{P}(h_y=r|x=x_c)$ is the probability that there are
$r-1$ points in the set of $h-1$ points in the ball around a `seed
point' that have the same label, $x_c$, as the seed. The $h-1$ points
in the ball are picked from the $n-1$ remaining data points and, under
the assumption that $X$ and $Y$ are independent, this is a random
choice. As such $\mathbb{P}(h_y=r|x=x_c)$ is the probability of
picking $r$ of the $n_c-1$ distinguished point, the remaining points with label
$x_c$, when selecting $h-1$ points from $n-1$. This means
\begin{equation}
    \mathbb{P} (h_y = r|x=x_c) =
    \frac{\binom{n_c-1}{r-1}\binom{n-n_c}{h-r}}{\binom{n-1}{h-1}} \sim
    \mbox{Hypergeometric}(n-1, n_c-1, h-1)
\end{equation}
Using $u(n-1,n_c-1,h-1)$ for the hypergeometric pdf, this means
\begin{equation}
  I_b=\sum^{h}_{r=1} \sum_{c=1}^{n_x}\frac{n_c}{n}u(n-1,n_c-1,h-1)\log_2  {\frac{n_c r}{h}}  
\end{equation}
and the unbiased estimator is
\begin{equation}
  I_e=I_0-I_b
\end{equation}

One subtlety concerns draws: what to do if there are multiple points
the same distance from the seed point. This can be dealt with by
counting points fractionally. Draws are only problematic if they occur
at the boundary of the ball around a point $(x_i,y_i)$, meaning that
it is not clear which $h$ points to regard as part of the
ball. Consider the case where there are $b$ points on the boundary, so
that $b$ points are equidistant from $(x_i,y_i)$, let $c<h$ be the
number of points closer to $(x_i,y_i)$ than the boundary points. This
means $b+c>h$, and only $h-c$ of the $b$ boundary points are needed to
fill out the ball. This is solved by counting all $b$ points
fractionally with a weight $(h-c)/b$; when calculating $h_y(i)$ any of
these $b$ points with the same label as $x_i$ adds this weight to the
total.

This leaves open the choice of $h$, the smoothing parameter; one approach that appears to work is
picking the $h$ that maximizes $I_e$ \cite{Houghton2019}.

\subsubsection*{Code availability}
Code to implement this estimator is available at \texttt{github.com/EstimatingInformation/DiscreteEstimator}.

\subsubsection*{Acknowledgements}
JW is support by an EPSRC Doctoral Training Programme scholarship.

\end{document}